\documentclass{PoS}

\usepackage{epsfig} 

\title{Study of the characteristics of SiPMs matrix as a photosensor for the scintillation detectors}

\ShortTitle{Study of the characteristics of SiPMs matrix ...}

\author{\speaker{I.M. Dzaparova}, A.M. Gangapshev, Yu.M. Gavrilyuk, V.B.~Petkov, A.V.~Sergeev, V.I.~Volchenko, S.P. Yakimenko, A.F. Yanin\\
        Institute for Nuclear Research of RAS\\
        E-mail: \email{dzaparova@yandex.ru}}

\abstract{The matrices formed of silicon photomultipliers (SiPMs) are very promising photosensors for the scintillation detectors. The use of SiPM matrices with appropriate optical collector gives, in principle, a possibility to do a snapshot of glowing track of charged particle traversing a scintillator. The prototype of such scintillation detector is under development now in INR RAS.
The preliminary results of characterization study of the matrix ArrayC-60035-64P-PCB (SensL company) for the prototype of such detector are presented.}

\FullConference{International Conference on New Photo-detectors - PhotoDet 2015\\
		6-9 July 2015\\
		Moscow, Troitsk, Russia}

\begin{document}

\section{Introduction}
Silicon photomultipliers are considered today in the world as a relevant replacement for traditional vacuum photomultipliers and are widely used in high energy physics, neutrino physics and cosmic ray physics \cite{Garutti}. The recent development of the technology has led to the emergence of matrices formed of SiPM. 
A charged particle traversing a scintillator induces scintillation along its track. At each point of the track the produced light is emitted isotropically. The use of SiPM matrices (with appropriate optical collector) gives, in principle, a possibility to do a snapshot of this glowing track \cite{Petkov15}. This technique has the obvious advantages. Firstly, the snapshot of glowing track of the particle gives a possibility to determine the direction of the  particle. Secondly, there is a possibility to measure the energy release along the track of particle. This technique is under development now in INR RAS. The SiPM matrices ArrayC-60035-64P-PCB are used as multi-pixel photodetectors in the prototype of such scintillation detector. The matrix ArrayC-60035-64P-PCB consists of 64 SiPMs (8$\times$8) SiPMs, each of them has size of 6$\times$6 mm$^2$ and consists of 18 980 micropixels \cite{Sensl}.

Use of this techniques implies that different SiPMs see the different parts of the particle track. Thereby for the correct track reconstruction the operation parameters for all individual SiPMs of the matrix must be known.
Study of extremely rare reactions and decays needs to diminish the background caused by the decay of natural radioactive elements always present in the facility. For such applications radiopurity of the matrices (and preamplifiers plates) must previously be measured.
 
\section{The number of photoelectrons vs. amplitude of pulses}
\begin{figure}[tph!]
\centerline{\psfig{file=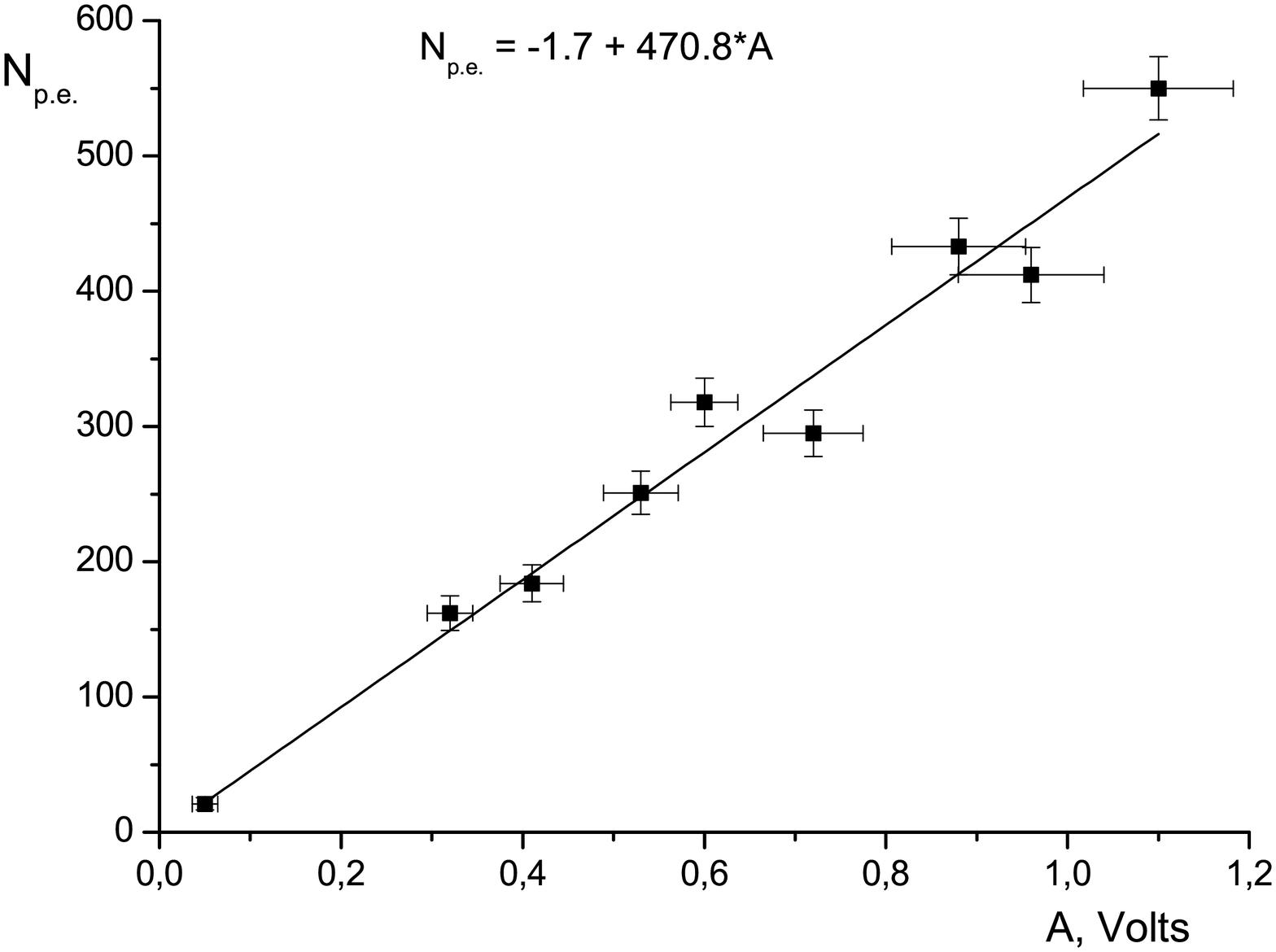,width=12cm}}
\caption{The number of photoelectrons vs. amplitude of SiMPs pulses.}
 \label{fig1}
\end{figure}

A conversion from the amplitude of SiPM pulse to the number of photoelectrons is needed for the track  analysis. The number of photoelectrons $N_{p.e.}$ was obtained by usual procedure using amplitude distribution of the SiPMs pulses at fixed luminous flux from LED:
\begin{equation}
N_{p.e.} =\left(   \frac{\bar{A}}{\sigma (A)} \right)^2,
\end{equation}
where $\bar{A}$ is mean value of pulses amplitude at fixed luminous flux and $\sigma (A)$ is a standard deviation of the amplitude distribution. The dependence of the number of photoelectrons on amplitude of pulses for one SiMP of the matrix is shown in Fig. \ref{fig1}.

\section{The scatter of amplification factors}
\begin{figure}[tph!]
\centerline{\psfig{file=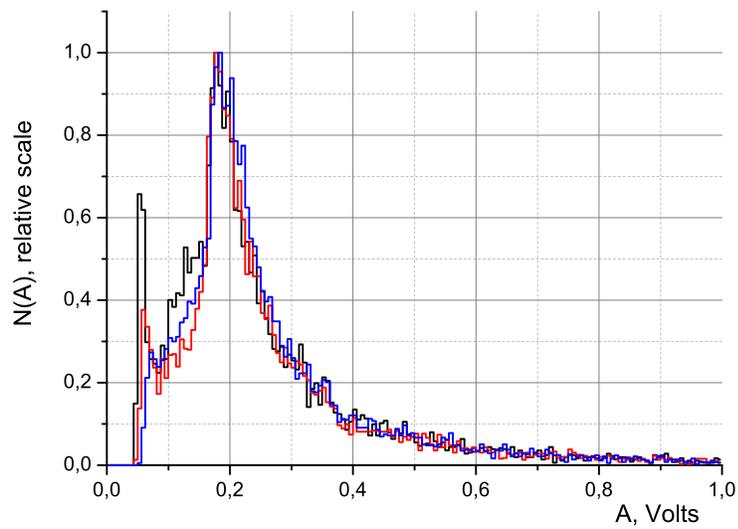,width=12cm}}
\caption{Normalized amplitude spectra for three SiPMs of the matrix.}
 \label{fig2}
\end{figure}

The scatter of amplification factors was estimated on scatter of the maxima position of the amplitude spectra of all individual SiPMs of the matrix. 
The measurements of the amplitude spectra of all SiPMs were carried out using the plate of plastic scintillator with sizes 500 mm $\times$ 500 mm and thickness of 50 mm. The matrix was placed in the center of the scintillator plate. The plate is much more than matrix, therefore the individual SiPMs are situated in the identical conditions from point of view of light collection. Normalized amplitude spectra for three SiPMs of the matrix are shown in Fig. \ref{fig2}. The distribution of the maxima position of the amplitude spectra of all SiPMs was obtained; the standard deviation of the distribution is 4.9\%.

\section{Quantum efficiency vs. angle  of incidence}
\begin{figure}[tph!] 
\centerline{\psfig{file=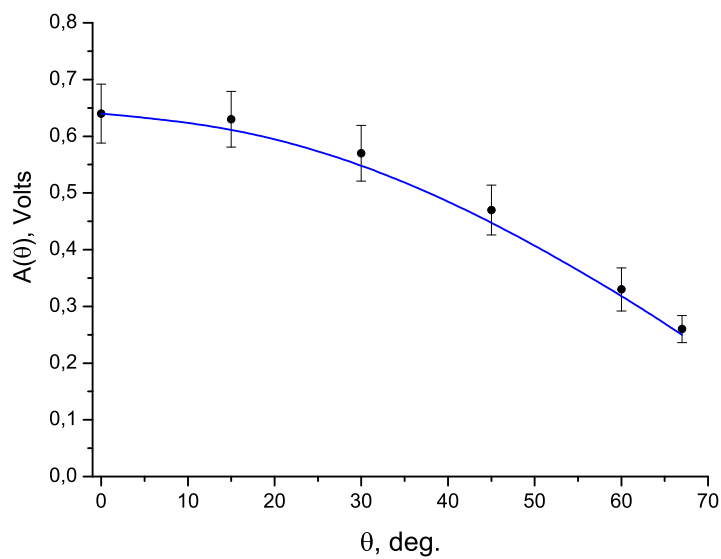,width=12cm}} 
\caption{Amplitude dependence of the pulse of the SiPM on the angle of incidence light beam. Points - experimental data, line - dependence $A(0)\times \cos \theta$.} \label{fig3}
\end{figure}

The effective area of a silicon photomultiplier depends on the angle of incidence $\theta$ of the light beam according to the law $\cos \theta$. 
If quantum efficiency of the silicon photomultiplier depends on the angle of incidence then one can expected that amplitude distribution of the SiPMs pulses will not be scaled according to the same law.
The experimental verification of this assumption has been performed. To create a parallel beam of uniform density the LED of blue glow has been removed from the matrix on 275 mm. The LED emitter was operated in the regime of low flux of light. In this regime only (1 - 1.5)\% of the total number of pixels of the SiPMT were simultaneously in operation.
Measuring the amplitude of pulses of the SiPM in response to the LED pulses was conducted for different angles of incidence photons with the same settings of the generator, i.e. at constant radiated power. For each angle 10000 pulses were collected, the results are presented in Fig. \ref{fig3}. The errors of measurements don't exceed 4.2\%. One can see that the amplitude of pulses of the SiPM depends on angle according to the law $\cos \theta$.

\section{Radiopurity of the matrix and preamplifiers plate}
The contamination of the matrix array and preamplifiers plate by gamma-active isotopes have been measured with use low background spectrometers at Baksan Neutrino Observatory of INR RAS \cite{Kuzminov}. NIKA and SNEG spectrometers were used for the measurements \cite{Busanov}. Results of these measurements are shown in Table 1 in comparison with corresponding measurements for copper, lead and PMT ET 9302B Enterprises.

One can see that the contamination of the matrix by gamma-active impurities is comparable with that for low background PMT ET 9302B Enterprises. 

\begin{table*}
\label{table:1}
\begin{tabular}{|c|c|c|c|c|c|}
\hline
Isotope, & \multicolumn{5}{|c|}{Activity of isotope, Bq/piece (Bq/kg)}\\
\cline{2-6}
gamma-line energy &  Matrix & Plate & Copper & Lead & PMT \\
\hline
$^{40}$K, & $0.052\pm 0.011$  & $0.038\pm 0.003$ & - & - & $0.26\pm 0.01$ \\
1460.8 keV & ($2.0\pm 0.4$) & ($7.1\pm 0.6$) & ($\le 2.3\cdot 10^{-3}$)  & ($\le 4.0\cdot 10^{-3}$) & ($1.71\pm 0.09$) \\
\hline
$^{208}$Tl, & $0.025\pm 0.003$  & $0.022\pm 0.001$ & - & -  & $(5.4\pm 0.8)\cdot 10^{-3}$  \\
2614.5 keV & $(1.0\pm 0.1)$  & $(4.1\pm 0.2)$  & $(\le 1.1\cdot 10^{-4})$ & $(\le 5.5\cdot 10^{-4})$         & ($0.036\pm 0.006$) \\
\hline
$^{214}$Bi, & $0.064\pm 0.004$  & $0.046\pm 0.001$ & - & -  & $0.180\pm 0.004)$  \\
609.3 keV & $(2.5\pm 0.1)$  &  $(8.6\pm 0.2)$  & $(\le 5.5\cdot 10^{-4})$ & ($0.0093\pm 0.0021$) & ($1.21\pm 0.03$) \\
\hline
$^{228}$Ac, & $0.083\pm 0.007$  & $0.076\pm 0.002$ & - & -  & $0.015\pm 0.00$  \\
911.2 keV & $(3.3\pm 0.3)$  &  $(14.2\pm 0.4)$  & ($\le 5.5\cdot 10^{-4}$) & ($\le 1.3\cdot 10^{-3}$) & ($0.1\pm 0.01$) \\
\hline
\end{tabular}\\[2pt]
\caption{Contamination of the matrix and the preamplifiers plate by the gamma-active isotopes. For comparison, the same values for copper, lead and PMT ET Enterprises 9302B are given.}
\end{table*}

\section{Conclusion}
Measurements of the operation parameters needed for the correct track reconstruction were carried out for all individual SiPMs of the matrix. 
The spread of the amplification factors of SiPMs has been measured; the standard deviation of the distribution is 4.9\%. The conversion from the amplitude of the pulses to the number of photoelectrons has been obtained. The quantum efficiency of the silicon photomultipliers don't depends on the angle of incidence of the light beam in range of 0-70 degrees within the uncertainty of measurements.
 
Measurements of the proper gamma background of the matrix have been performed. The contamination of the matrix by gamma-active impurities is low enough and 
comparable with that for low background PMT ET 9302B Enterprises. 

\vspace{5mm}

{\bf Acknowledgements.}\\
This work was supported by the the Russian Foundation for Basic Research (grant 14-22-03075).

\end{document}